\documentclass[superscriptaddress,aps,prl,showkeys,showpacs,twocolumn,10pt]{revtex4-1}
\usepackage[OT1,T1]{fontenc}
\usepackage{graphicx}
\usepackage{rotating}
\begin{document}

\title{Evidence for a New Resonance from Polarized Neutron-Proton Scattering}
\date{\today}

\newcommand*{\IKPUU}{Division of Nuclear Physics, Department of Physics and 
 Astronomy, Uppsala University, Box 516, 75120 Uppsala, Sweden}
\newcommand*{\ASWarsN}{Department of Nuclear Physics, National Centre for 
 Nuclear Research, ul.\ Hoza~69, 00-681, Warsaw, Poland}
\newcommand*{\IPJ}{Institute of Physics, Jagiellonian University, ul.\ 
 Reymonta~4, 30-059 Krak\'{o}w, Poland}
\newcommand*{\PITue}{Physikalisches Institut, Eberhard--Karls--Universit\"at 
 T\"ubingen, Auf der Morgenstelle~14, 72076 T\"ubingen, Germany}
\newcommand*{\Kepler}{Kepler Center for Astro and Particle Physics, University
 of T\"ubingen, Auf der Morgenstelle~14, 72076 T\"ubingen, Germany}
\newcommand*{\MS}{Institut f\"ur Kernphysik, Westf\"alische 
 Wilhelms--Universit\"at M\"unster, Wilhelm--Klemm--Str.~9, 48149 M\"unster, 
 Germany}
\newcommand*{\ASWarsH}{High Energy Physics Department, National Centre for 
 Nuclear Research, ul.\ Hoza~69, 00-681, Warsaw, Poland}
\newcommand*{\IITB}{Department of Physics, Indian Institute of Technology 
 Bombay, Powai, Mumbai--400076, Maharashtra, India}
\newcommand*{\PGI}{Peter Gr\"unberg Institut, Forschungszentrum J\"ulich,
  52425 J\"ulich, Germany}
\newcommand*{\ILP}{Institut f\"ur Laser- und Plasmaphysik, Heinrich-Heine
  Universit\"at D\"usseldorf, 40225 D\"usseldorf, Germany}
\newcommand*{\IKPJ}{Institut f\"ur Kernphysik, Forschungszentrum J\"ulich, 
 52425 J\"ulich, Germany}
\newcommand*{\JCHP}{J\"ulich Center for Hadron Physics, Forschungszentrum 
 J\"ulich, 52425 J\"ulich, Germany}
\newcommand*{\Bochum}{Institut f\"ur Experimentalphysik I, Ruhr--Universit\"at 
 Bochum, Universit\"atsstr.~150, 44780 Bochum, Germany}
\newcommand*{\ZELJ}{Zentralinstitut f\"ur Engineering, Elektronik und 
 Analytik, Forschungszentrum J\"ulich, 52425 J\"ulich, Germany}
\newcommand*{\Erl}{Physikalisches Institut, 
 Friedrich--Alexander--Universit\"at Erlangen--N\"urnberg, 
 Erwin--Rommel-Str.~1, 91058 Erlangen, Germany}
\newcommand*{\ITEP}{Institute for Theoretical and Experimental Physics, State 
 Scientific Center of the Russian Federation, Bolshaya Cheremushkinskaya~25, 
 117218 Moscow, Russia}
\newcommand*{\Giess}{II.\ Physikalisches Institut, 
 Justus--Liebig--Universit\"at Gie{\ss}en, Heinrich--Buff--Ring~16, 35392 
 Giessen, Germany}
\newcommand*{\IITI}{Department of Physics, Indian Institute of Technology 
 Indore, Khandwa Road, Indore--452017, Madhya Pradesh, India}
\newcommand*{\Aachen}{III.~Physikalisches Institut~B, Physikzentrum, 
 RWTH Aachen, 52056 Aachen, Germany}
\newcommand*{\HepGat}{High Energy Physics Division, Petersburg Nuclear Physics 
 Institute, Orlova Rosha~2, Gatchina, Leningrad district 188300, Russia}
\newcommand*{\HeJINR}{Veksler and Baldin Laboratory of High Energiy Physics, 
 Joint Institute for Nuclear Physics, Joliot--Curie~6, 141980 Dubna, Russia}
\newcommand*{\Katow}{August Che{\l}kowski Institute of Physics, University of 
 Silesia, Uniwersytecka~4, 40-007, Katowice, Poland}
\newcommand*{\IFJ}{The Henryk Niewodnicza{\'n}ski Institute of Nuclear 
 Physics, Polish Academy of Sciences, 152~Radzikowskiego St, 31-342 
 Krak\'{o}w, Poland}
\newcommand*{\NuJINR}{Dzhelepov Laboratory of Nuclear Problems, Joint 
 Institute for Nuclear Physics, Joliot--Curie~6, 141980 Dubna, Russia}
\newcommand*{\KEK}{High Energy Accelerator Research Organisation KEK, Tsukuba, 
 Ibaraki 305--0801, Japan} 
\newcommand*{\ASLodz}{Department of Cosmic Ray Physics, National Centre for 
 Nuclear Research, ul.\ Uniwersytecka~5, 90--950 {\L}\'{o}d\'{z}, Poland}
\author{P.~Adlarson}    \affiliation{\IKPUU}
\author{W.~Augustyniak} \affiliation{\ASWarsN}
\author{W.~Bardan}      \affiliation{\IPJ}
\author{M.~Bashkanov}   \affiliation{\PITue}\affiliation{\Kepler}
\author{F.S.~Bergmann}  \affiliation{\MS}
\author{M.~Ber{\l}owski}\affiliation{\ASWarsH}
\author{H.~Bhatt}       \affiliation{\IITB}
\author{M.~B\"uscher}\affiliation
{\PGI}\affiliation{\ILP}
\author{H.~Cal\'{e}n}   \affiliation{\IKPUU}
\author{I.~Ciepa{\l}}   \affiliation{\IPJ}
% ...

\newcommand*{\Delhi}{Department of Physics and Astrophysics, University of 
 Delhi, Delhi--110007, India}
\newcommand*{\Mainz}{Institut f\"ur Kernphysik, Johannes 
 Gutenberg--Universit\"at Mainz, Johann--Joachim--Becher Weg~45, 55128 Mainz, 
 Germany}
\newcommand*{\Bern}{Albert Einstein Center for Fundamental Physics, University 
 of Bern, Sidlerstrasse~5, 3012 Bern, Switzerland}
\newcommand*{\Sheff}{Department of Physics and Astronomy, University of 
 Sheffield, Hounsfield Road, Sheffield, S3 7RH, United Kingdom}

\collaboration{WASA-at-COSY Collaboration}\noaffiliation

\newcommand*{\GW}{Data Analysis Center at the Institute for Nuclear Studies,
  Department of Physics, The George Washington University, Washington,
  D.C. 20052, U.S.A.}
\author{R. L. Workman}     \affiliation{\GW}
\author{W. J. Briscoe}     \affiliation{\GW}
\author{I. I. Strakovsky}  \affiliation{\GW}
\collaboration{SAID Data Analysis Center}

\begin{abstract}
Exclusive and kinematically complete high-statistics measurements of quasifree
polarized $\vec{n}p$ scattering have been performed in the energy region of the
narrow resonance-like structure $d^*$ with $I(J^P) = 0(3^+)$,
$M~\approx$~2380~MeV and $\Gamma \approx$ 70 MeV observed recently in the
double-pionic fusion channels $pn \to d\pi^0\pi^0$ and $pn \to d\pi^+\pi^-$.  
The experiment was carried out with the WASA detector setup at COSY having a
polarized deuteron beam impinged on the hydrogen pellet target and utilizing
the quasifree process $\vec{d}p \to np + p_{spectator}$. This allowed the $np$
analyzing power, $A_y$, to be measured over a broad angular range. The
obtained $A_y$ angular distributions deviate systematically from the current
SAID SP07 NN partial-wave solution. Incorporating the new $A_y$ data into the
SAID analysis produces a pole in the $^3D_3 - ^3G_3$ waves in support of the
$d^*$ resonance hypothesis. 
\end{abstract}


\maketitle

\section{Introduction}

Recent exclusive and kinematically complete measurements of the basic
double-pionic fusion reactions $pn \to d \pi^0\pi^0$ and $pn \to d \pi^+\pi^-$
revealed a narrow  resonance-like structure in the total cross section 
\cite{mb,MB,isofus} at a mass $M \approx$ 2380~MeV with a width of 
$\Gamma \approx$ 70 MeV, which is consistent with a $I(J^P) = 0(3^+)$ assignment
 \cite{MB}. Additional evidence for this structure has recently been found in
 the $pn \to pp\pi^0\pi^-$ reaction \cite{TS}, where it was denoted by  
$d^*$, following the notation associated with the so-called "inevitable
dibaryon" \cite{goldman}.

If the observed resonance-like structure truly constitutes an $s$-channel 
resonance in the neutron-proton system, then it must be seen in the 
observables of elastic $np$ scattering. In Ref. \cite{PBC} this
resonance effect in $np$ scattering has been estimated. There it was shown 
that a noticeable effect should appear in the analyzing power $A_y$, since
this observable is composed only of interference terms between partial waves,
thus being most sensitive to small changes in the partial waves.

For the analyzing power, there exist data only below and above the resonance
region. These data sets, at $T_n$~=~1.095 GeV ($\sqrt s$ = 2.36 GeV)
\cite{ball,les} and $T_n$~=~1.27 GeV ($\sqrt s$ = 2.43 GeV) \cite{mak,dieb},
exhibit very similar angular distributions. This gap in the existing
measurements of $A_y$ has motivated the present study.

\section{Experiment}

We have measured the energy dependence of polarized $\vec{n}p$ elastic
scattering in the quasifree mode. The experiment was carried out with the WASA
detector \cite{CB,wasa} at COSY (FZ J\"ulich), using a polarized deuteron beam
with an energy of $T_d$~=~2.27~GeV impinging on the WASA hydrogen pellet
target. With this setup, a full energy coverage of the conjectured resonance was
obtained. Note that we observe here the quasi-free scattering process 
$\vec d p \to np + p_{spectator}$ in inverse kinematics, which allows a
detection of the fast spectator proton in the forward detector of WASA.

Since we deal here with events originating from channels with large cross
section, the trigger was solely requesting one hit in the first layer of the
forward range hodoscope. This hit could originate from either a charged
particle or a neutron. For the case of quasifree $np$ scattering, this defines
three event classes, each having the spectator proton appearing in the 
forward detector:
\begin{itemize}
\item  scattered proton and scattered neutron both detected in the central
  detector, covering the neutron angle region $31^\circ < \Theta_n^{cm} <
  129^\circ$, 
\item scattered proton detected in the forward detector, with the scattered
  neutron being unmeasured, covering $132^\circ < \Theta_n^{cm} < 178^\circ$
  and 
\item scattered proton detected in the central detector, with the neutron being
  unmeasured, covering the angular range $30^\circ < \Theta_n^{cm} <
  41^\circ$.
\end {itemize}

Combining events, nearly the full range of neutron scattering angles could be 
covered.

Since, through the use of the inverse kinematics, the spectator proton is in
the beam particle, the deuteron, the spectator is very fast. This allows its
detection in the forward detector. By reconstruction of its kinetic energy and
its direction the full four-momentum of the spectator proton has been
determined. 

Similarly, the four-momentum of the actively scattered proton has been
obtained from its track information in either the forward or central detector
(in the latter case the energy information was not retrieved). 

Therefore, we have reconstructed the full event, including the four-momentum
of the unmeasured neutron, and even have one overconstraint in the subsequent
kinematic fit, when the neutron has not been measured explicitly. 

In the case where the neutron has been detected by a hit in the calorimeter
(composed of 1012 CsI(Na) crystals) of the central detector -- associated with
no hit in the preceding plastic scintillator barrel, the directional
information of the scattered neutron has also been obtained. Therefore, these
events have undergone a kinematic fit with two overconstraints. 

In order to avoid a distortion of the beam polarization, the magnetic field of
the solenoid in the central detector was switched off. The measurements were
carried out with cycles of the beam polarization "up", "down" and unpolarized
(originating from the same polarized source), where "up" and  "down" refers to
a horizontal scattering plane. We verified that the beam, originating from the
polarized source, indeed was unpolarized when using it in its "unpolarized"
mode. This was accomplished by comparing the azimuthal angular dependence of
the scattered events to that obtained through the use of a conventional
unpolarized source. 

The magnitude of the beam polarization was determined and monitored by $dp$
elastic scattering, which was measured in parallel by detecting the scattered
deuteron in the forward detector as well as the associated scattered proton in
the central detector. The vector and tensor components of the deuteron beam
were obtained by fitting our results, for the vector and tensor analyzing
power, to those obtained previously at ANL \cite{ANL} for $T_d$ = 2.0 GeV and
more recently at COSY-ANKE \cite{ANKE} at $T_d$ = 2.27 GeV. As a result we
obtained beam polarizations of $P_z = 0.67(2)$, $P_{zz}$ = 0.65(2) for "up"
and $P_z$ = -0.45(2), $P_{zz}$ = 0.17(2) for "down". 
The vector polarization of the beam, for quasifree scattering, has been
checked by quasifree $pp$ scattering. This was also measured in parallel by
detecting one of the protons in the forward detector and the other one in the
central detector, in addition, checking their angular correlation for elastic
events. Our results for the quasifree $pp$ analyzing power are in quantitative
agreement both with the EDDA measurements \cite{EDDA} of free $pp$ scattering
and with the current SAID phase shift solution SP07 \cite{Arndt07}. 

Since we have measurements with spin "up", "down" and unpolarized, the vector
analyzing power can be derived in three different ways, by using each two of
the three spin orientations. All three methods should give identical
results. Differences may be taken as an estimate of systematic uncertainties
which are added quadratically to the statistical ones to give the total
uncertainties plotted in Figs.~1,2 and 4. 

The momentum distribution of the observed spectator proton, in the elastic
$np$ scattering process, agrees with Monte Carlo simulations of the proton
momentum distribution in the deuteron filtered by the acceptance of the WASA
detector. 
In order to assure a
quasi-free process, we omit events with spectator momenta larger than 0.16
GeV/c (in the deuteron rest system) as done in previous work
\cite{MB,isofus}.

\section{Results and Discussion}

Due to the Fermi motion of the nucleons bound in the beam deuteron, the
measurement of the quasi-free $np$ scattering process covers a range of
energies in the $np$ system. Meaningful statistics could be collected for the
range of $np$ center-of-mass energies 2.36 $< \sqrt s <$ 2.41 GeV
corresponding to $T_n$ = 1.10 - 1.20 GeV. First, we show the data (solid
circles) in Fig.~1 without selecting specific $np$ center-of-mass energies,
{\it i.e.} without accounting for the spectator momentum. Hence this data set
corresponds to the weighted average over the covered interval of $\sqrt
s$. The solid line represents the current SAID SP07 partial-wave solution
\cite{Arndt07}, whereas dashed and dotted lines give the results of revised
SAID partial-wave analyses, including the WASA dataset, as described below.
Next, we have taken the measured spectator four-momentum into account and
constructed the effective $\sqrt s$ for each event. We thus obtained angular
distributions sorted into six $\sqrt s$ bins, two of which are shown in Fig.~2
as examples. All of our data deviate strikingly from the SP07 solution.

\begin{figure} 
\centering
\includegraphics[width=0.99\columnwidth]{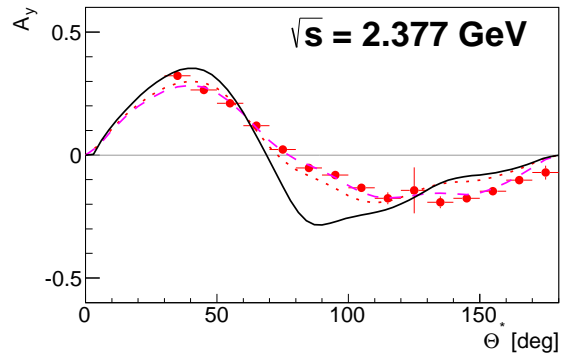}
\caption{\small (Color online) Angular distribution of the $np$ analyzing
  power without consideration of the spectator momentum, 
  corresponding to a weighted average over the measured interval $\sqrt s$ =
  2.367 - 2.403 GeV ($T_n$ = 1.108 - 1.197 GeV) with a centroid at
  $\sqrt s$ = 2.377 GeV.
  The results from this work are shown as solid circles
  with error bars including both statistical and systematic uncertainties. The
  solid line represents the SAID SP07 phase shift prediction \cite{Arndt07},
  whereas the dashed (dotted) line gives the result of the new weighted
  (unweighted) SAID partial-wave solution (see text).  
}
\label{fig1}
\end{figure}

\begin{figure}
\centering
\includegraphics[width=0.99\columnwidth]{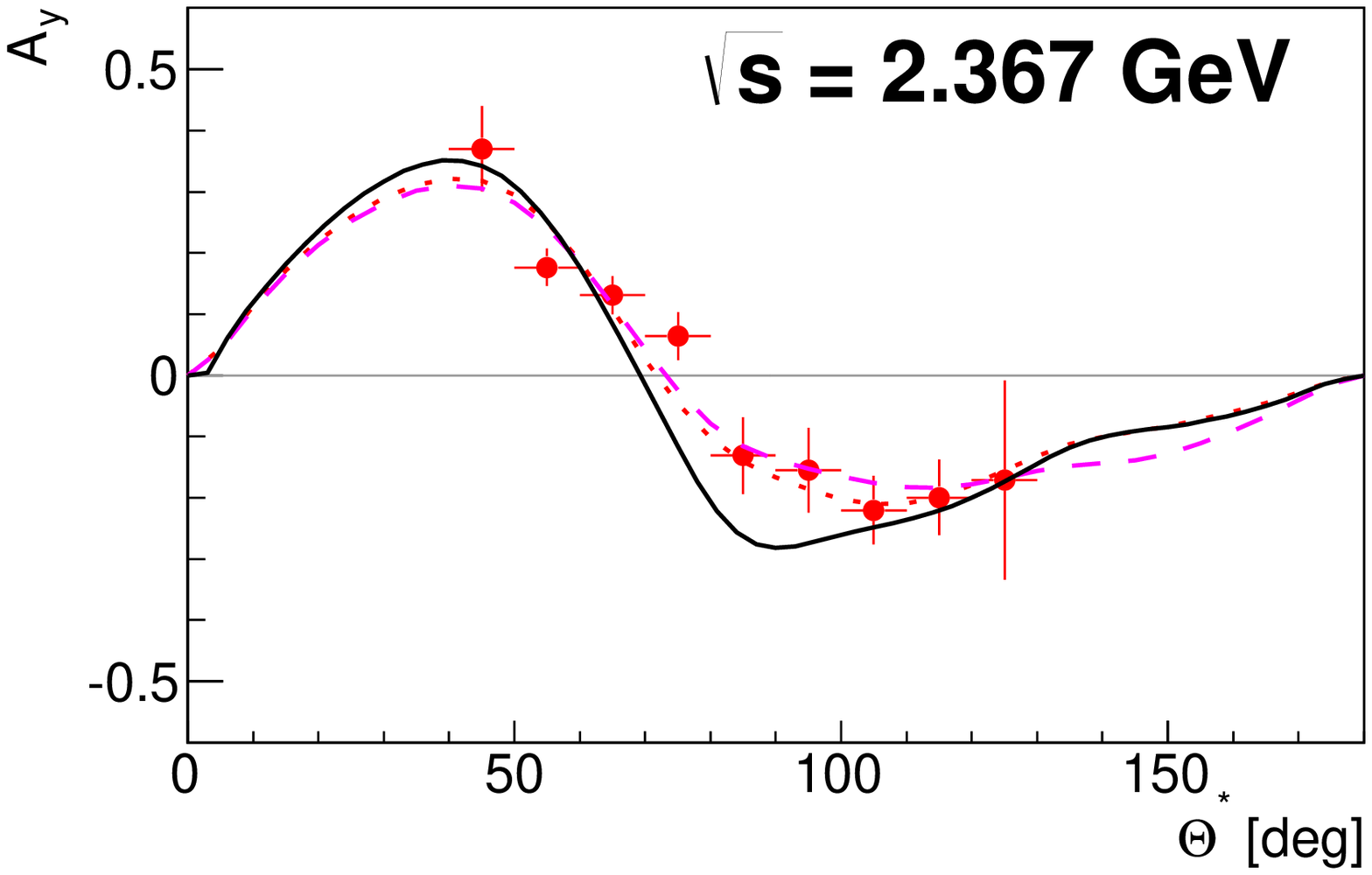}
\includegraphics[width=0.99\columnwidth]{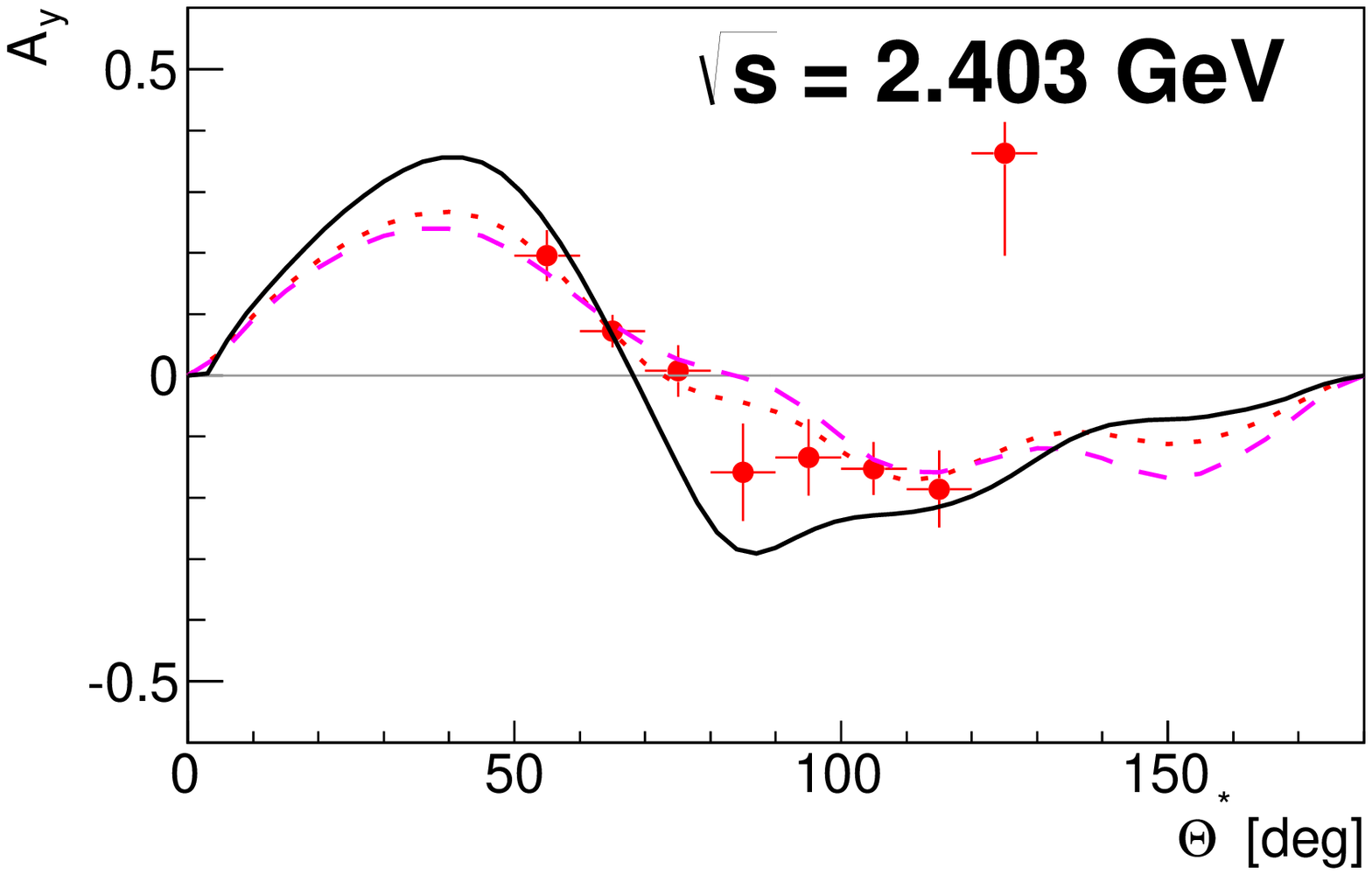}
\caption{\small (Color online) Notation as in Fig.~1, but for  $\sqrt s$ =
  2.367 (top) and 2.403 GeV (bottom) corresponding to $T_n$ = 1.11 and 1.20
  GeV. The full symbols denote results from this work taking into account the
  spectator four-momentum information. For the meaning of the curves see Fig.~1.
}
\label{fig2}
\end{figure}


As a test, the present $A_y$ data set was included in the SAID database and the
phenomenological approach used in generating the $NN$ partial-wave solution,
SP07~\cite{Arndt07}, was retained. Here we first considered whether the
existing form was capable of describing these new $A_y$ measurements. One
advantage of this approach is that the employed Chew-Mandelstam K-matrix can
produce a pole in the complex energy plane without the explicit inclusion of a
K-matrix pole in the fit form. Neither the existence of a pole nor the
effected partial waves are predetermined. A detailed overview of this
formalism is given in Ref.\cite{Arndt87}.



The fitted $A_y$ data were angular distributions at $T_{\rm Lab}$ values of
1.108, 1.125, 1.135, 1.139, 1.156, 1.171, and 1.197 GeV.  A first attempt to
fit this dataset started from the functional form of the current SP07 fit, and
only varied the associated free parameters. A $\chi^2$/datum of 1.8 was found
for all angular distributions, apart from the one at 1135 MeV. This was fairly
consistent with the overall $\chi^2$/datum given by the global fit of $np$
elastic scattering data to 2 GeV. However, the set at 1135 MeV contributed a
$\chi^2$/datum of about 25, having better statistics and a wider angular
coverage.  

The fit parameters are expansion coefficients for the K-matrix elements, which
are smooth in energy; either polynomials or basis elements having required
left-hand cuts, as described in Ref.\cite{Arndt87}. Failing to reproduce the
1135 MeV set, the fit form was scanned to find partial waves for which an
added term in the K-matrix expansion produced the most efficient reduction in
$\chi^2$. The addition of parameters and re-fitting resulted in a rapid
variation of the coupled $^3D_3$ and $^3G_3$ waves in the vicinity of the
problematic 1135 MeV data set. 

Some weighting was necessary in this fit, as only a few angular points from
the full set were determining the altered energy dependence.  The fit was
repeated with different weightings for the 1135 MeV $A_y$ dataset. Having
found a better fit at 1135 MeV, a subsequent fit was produced without
weighting. These, qualitatively similar, results are compared in the figures.





In Fig.~1 we plot the fit to the 1135 MeV angular distribution from the SP07
prediction (not including the new data), a weighted fit ( errors decreased by
a factor of 4 ), and an unweighted fit including the present dataset and using
the new fit form. 

Resulting changes in the $^3D_3$-$^3G_3$ coupled waves are displayed in
Fig.~3. Here the $^3D_3$ wave obtained a typical resonance-like shape, whereas
the $^3G_3$ wave changed less dramatically. A search of the complex energy
plane revealed a pole in the coupled $^3D_3$-$^3G_3$ wave. Other partial waves
did not change significantly over the energy range spanned by the new
data. Fig.~3 also displays single-energy solutions, generated from the old
SP07 fit. These discrete points are fits to data within narrow energy bins,
allowing amplitude variations to produce a best fit to data, and are used to
search for systematic deviations from the global fit~\cite{Arndt87}. In the
$^3D_3$ partial-wave plot near 1135 MeV, the new fit appears to agree with
these single-energy results much better than SP07.

The fit repeated with different weightings for the new $A_y$ data resulted in
a variation of the pole position and could be considered a minimal 'error' on
its value within the present fit form. In the weighted fits, a pole was
located at (2392 - i37) MeV. The re-fit without weighting produced a pole with 
(2385 - i39) MeV. Together with a speed-plot determination we arrive at
($2380\pm10 - i 40\pm5$) MeV as our best estimate for the pole position. 

\begin{figure}
\centering
\includegraphics[width=0.89\columnwidth]{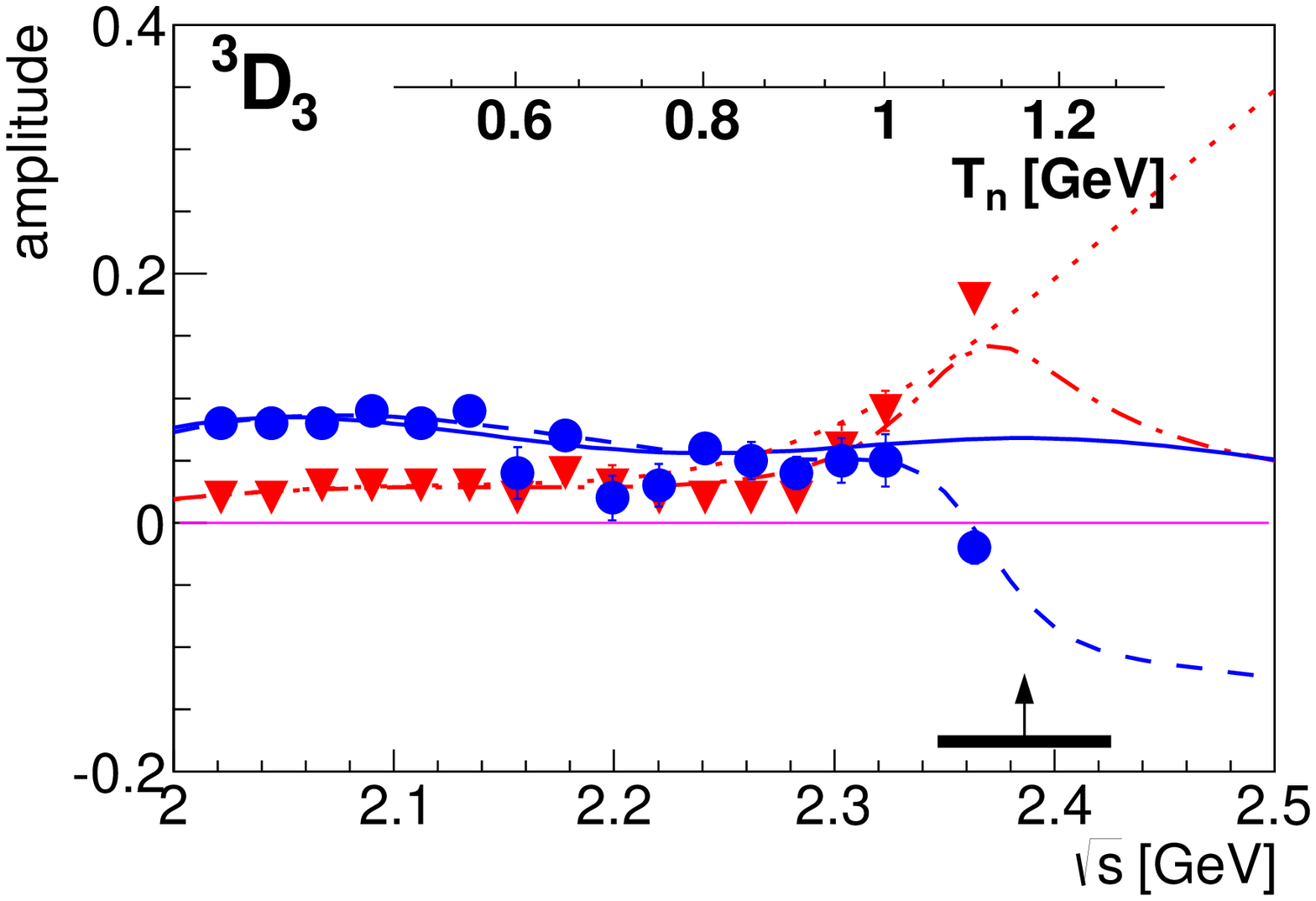}
\includegraphics[width=0.89\columnwidth]{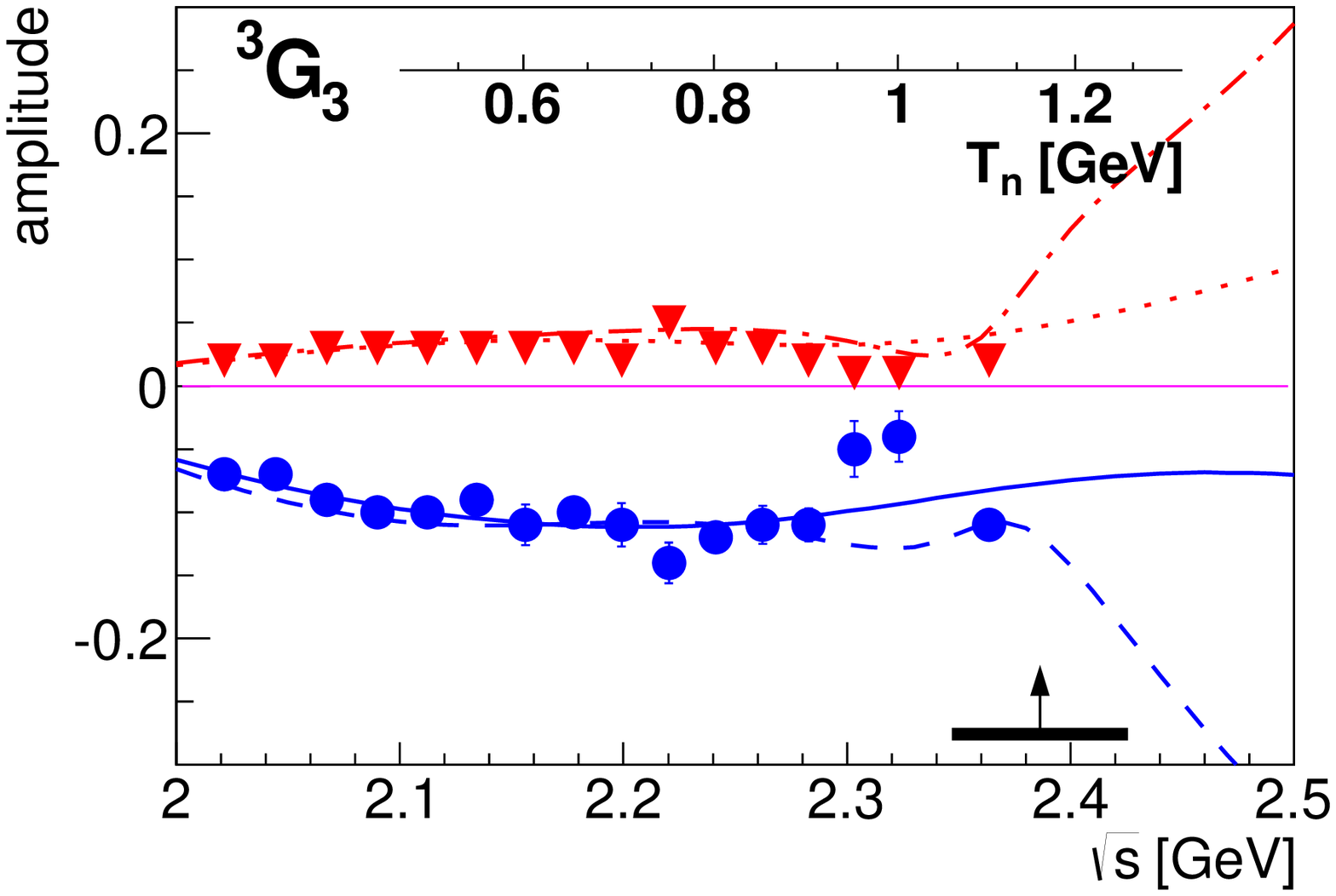}
\includegraphics[width=0.89\columnwidth]{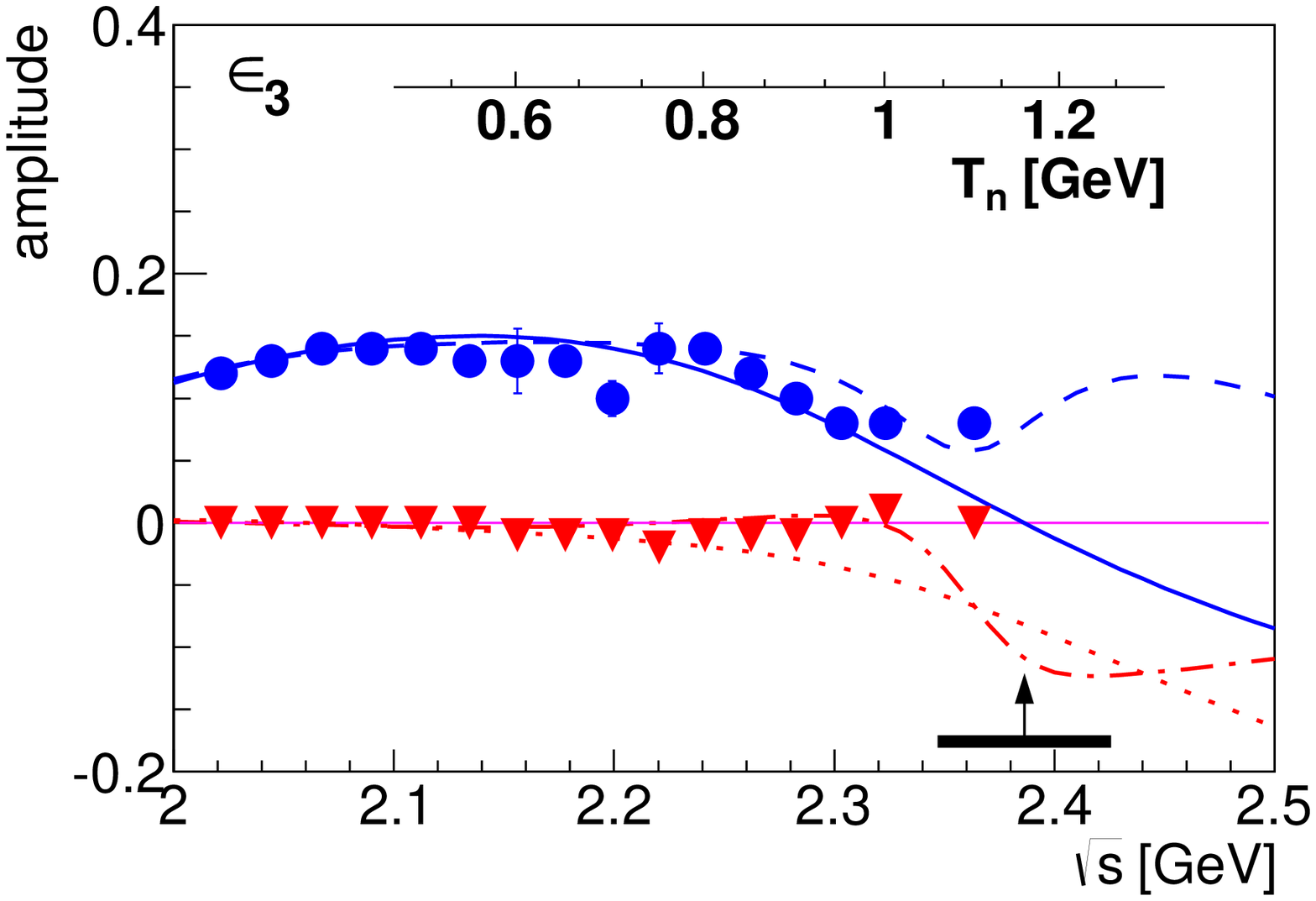}
\caption{(Color online) Changes to the (dimensionless) $^3D_3$ (top) and
  $^3G_3$ (middle) partial waves including their mixing amplitude $\epsilon_3$
  (bottom). Solid (dotted) curves give the real (imaginary) part of the
  partial-wave amplitudes from SP07, whereas  the dashed (dash-dotted) curves
  represent the new (weighted) solution. Results from previous single-energy
  fits \cite{Arndt07} are shown as solid circles (real part) and inverted
  triangles (imaginary part). Vertical arrows and horizontal bars indicate mass
  and width of the resonance (estimated from the pole position).
}
\label{fig:f2}
\end{figure}


\begin{figure}
\centering
\includegraphics[width=0.99\columnwidth]{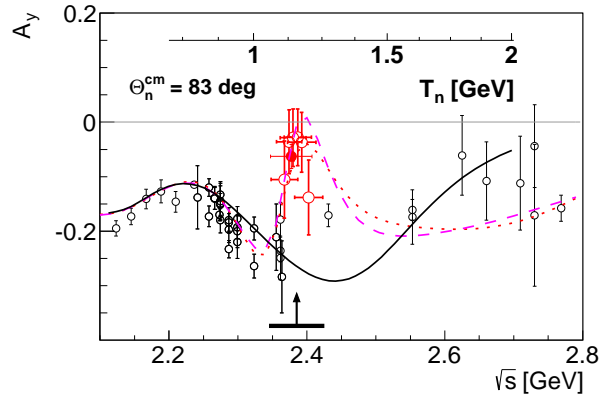}
\caption{\small (Color online) Energy dependence of the $np$ analyzing power
  at $\Theta_n^{cm}$ = 83$^\circ$. The solid symbols denote the results from
  this work, the open symbols those from previous work
  \cite{ball,les,new,arn,bal1,mcn,gla,mak}. For the meaning of the curves see
  Fig.~1. Vertical arrow and horizontal bar indicate mass and width of the
  resonance (estimated from the pole position). 
}
\label{fig4}
\end{figure}

From the decomposition of the $np$ observables into partial-wave amplitudes
\cite{ArndtRoper}, it follows that the  resonance contribution in $A_y$ is
proportional to the associated Legendre polynomial
$P_3^1(cos\Theta_n^{cm})$. $P_3^1$ is maximal at $\Theta_n^{cm} = 31.1^\circ$
and minimal at $90^\circ$. Since at the latter angle the differential cross
section is at minimum and much lower than at the former angle, the resonance
effect in $A_y$ becomes maximal at $\Theta_n^{cm} = 90^\circ$. To check this
behavior, we have inspected the energy dependence of $A_y$. In order to 
include a reasonable number of previous measurements, we have chosen 
the nearby angle $\Theta_n^{cm} = 83^\circ \pm 2^\circ$ to be plotted in
Fig.~4. The data exhibit a pronounced resonance-like behavior in accordance
with the new partial-wave solution -- in tendency even somewhat narrower.

\section{Summary and Conclusions}

In conclusion, our exclusive and kinematically complete measurement of
quasi-free polarized $\vec{n}p$ scattering provides detailed high-statistics data
for the analyzing power in the energy range, where previously a narrow
resonance-like structure with $I(J^P) = 0(3^+)$ was observed in the
double-pionic fusion to deuterium. A partial-wave analysis including the new
$np$ scattering data exhibits a resonance pole in the coupled $^3D_3$ -
$^3G_3$ partial waves in accordance with the expectation of a $d^*$ resonance
structure. This structure has been associated with a bound $\Delta \Delta$
resonance, which could contain a mixture of asymptotic $\Delta \Delta$
\cite{dyson} and 6-quark, hidden color, configurations~\cite{Brodsky}. Though
less exotic explanations
cannot be excluded at the present stage, dibaryon systems matching the mass
and width of this dibaryon candidate have been recently successfully generated
within 3-body~\cite{Gal} and quark model~\cite{qm} calculations. It should be
noted that earlier dibaryon candidates~\cite{Arndt87} were widely discounted
due to their appearance near the $N\Delta$ cut and the possibility of a
pseudo-resonance mimicking their behavior. Such complications do not arise
here -- though we note the existence of a nearby $NN^*(1440)$
threshold. However, we are not aware of any mechanism by which the very broad
Roper resonance could induce the narrow resonance structure considered here.

Finally, we note that the new partial-wave solution improves also the
description of total cross section data as well as polarization observables
obtained at ANKE \cite{ANKEAXX} in the resonance region. A full account of the
new results will be given in an extended forthcoming paper.


We acknowledge valuable discussions with J. Haidenbauer, C. Hanhart,
A. Kacharava and C. Wilkin on this issue. This work has been supported by
BMBF, Forschungszentrum J\"ulich (COSY-FFE), the U.S. Department of Energy 
Grant DE-FG02-99ER41110, the Polish National Science Centre (grant
No. 2011/03/B/ST2/01847) and the Foundation for Polish Science (MPD).

\end{document}